\documentclass{llncs}

\usepackage{paradise}
\usepackage{longtable,booktabs}
\usepackage{graphicx}
\usepackage[sectionbib]{natbib}
\setcitestyle{numbers,square,comma}

\newcommand{\hypertarget}[2]{#2}

\usepackage{caption}
\begin{document}
\bibliographystyle{plainnat}

\pagestyle{plain}
\mainmatter

\title{Reproducible Execution of \posix{} Programs with \dios{}\thanks{This work has been partially supported by the Czech Science Foundation grant No.~18-02177S and by Red Hat, Inc.}}

\author{Petr Ročkai \and Zuzana Baranová \and Jan Mrázek \and \break Katarína Kejstová \and Jiří Barnat}
\institute{\fimuni\\ \{xrockai,xbaranov,xmrazek7,xkejstov,barnat\}@fi.muni.cz}

\maketitle

\begin{abstract}
  In this paper, we describe \dios{}, a lightweight model operating system which can be used to execute programs that make use of \posix{} APIs. Such executions are fully reproducible: running the same
  program with the same inputs twice will result in two exactly identical instruction traces, even if the program uses threads for parallelism.
  
  \dios{} is implemented almost entirely in portable C and C++: although its primary platform is \divm{}, a verification-oriented virtual machine, it can be configured to also run in \klee{}, a symbolic
  executor. Finally, it can be compiled into machine code to serve as a user-mode kernel.
  
  Additionally, \dios{} is modular and extensible. Its various components can be combined to match both the capabilities of the underlying platform and to provide services required by a particular program.
  New components can be added to cover additional system calls or APIs.
  
  The experimental evaluation has two parts. \dios{} is first evaluated as a component of a program verification platform based on \divm{}. In the second part, we consider its portability and modularity by
  combining it with the symbolic executor \klee{}.
\end{abstract}

\hypertarget{introduction}{%
\section{Introduction}\label{introduction}}

Real-world software has a strong tendency to interact with its execution environment in complex ways. To make matters worse, typical environments in which programs execute are often extremely
unpredictable and hard to control. This is an important factor that contributes to high costs of software validation and verification. Even the most resilient verification methods (those based on
testing) see substantial adverse effect.

In automated testing, one of the major criteria for a good test case is that it gives reliable and reproducible results, without intermittent failures. This is especially true in the process of
debugging: isolating a fault is much harder when it cannot be consistently observed. For this reason, significant part of the effort involved in testing is spent on controlling the influence of the
environment on the execution of test cases.

The situation is even worse with more rigorous verification methods -- for instance, soundness of verification tools based on dynamic analysis strongly depends on the ability to fully control the
execution of the system under test.

In this paper, we set out to design and implement a small and sufficiently self-contained model operating system that can provide a realistic environment for executing \posix{}-based programs. Since this
environment is fully virtualised and isolated from the host system, program execution is always fully reproducible. As outlined above, such reproducibility is valuable, sometimes even essential, in
testing and program analysis scenarios. Especially dynamic techniques, like software model checking or symbolic execution, rely on the ability to replay interactions of the program and obtain
identical outcomes every time.

\hypertarget{contribution}{%
\subsection{Contribution}\label{contribution}}

The paper describes our effort to implement a~compact operating system on top of existing verification frameworks and virtual machines (see Section~\ref{sec:supported}). Despite its minimalist design,
the current implementation covers a wide range of \posix{} APIs in satisfactory detail (see also Section~\ref{sec:syscalls}). The complete source code is available online\footnote{\url{https://divine.fi.muni.cz/2019/dios/}},
under a permissive open-source licence. Additionally, we have identified a set of low-level interfaces (see Section~\ref{sec:platform}) with two important qualities:

\begin{enumerate}
\def\labelenumi{\arabic{enumi}.}
\tightlist
\item
  the interfaces are lightweight and easy to implement in a VM,
\item
  they enable an efficient implementation of complex high-level constructs.
\end{enumerate}

Minimal interfaces are a sound design principle and lead to improved modularity and re-usability of components. In our case, identification of the correct interfaces drives both \emph{portability} and
\emph{compactness of implementation}.

Finally, the design that we propose improves robustness of verification tools. A common implementation strategy treats high-level constructs (e.g.~the \texttt{pthread} API) as primitives built into
the execution engine. This ad-hoc approach often leads to implementation bugs which then compromise the soundness of the entire tool. Our design, on the other hand, emphasises clean separation of
concerns and successfully reduces the amount of code which forms the trusted execution and/or verification core.

\hypertarget{sec:goals}{%
\subsection{Design Goals}\label{sec:goals}}

We would like our system to have the following properties:

\begin{enumerate}
\def\labelenumi{\arabic{enumi}.}
\tightlist
\item
  Modularity: minimise the interdependence of the individual OS components. It should be as easy as possible to use individual components (for instance \texttt{libc}) without the others. The kernel
  should likewise be modular.
\item
  Portability: reduce the coupling to the underlying platform (verification engine), making the OS useful as a pre-made component in building verification and testing tools.
\item
  Veracity: the system should precisely follow \posix{} and other applicable standardised semantics. It should be possible to port realistic programs to run on the operating system with minimal effort.
\end{enumerate}

Since the desired properties are hard to quantify, we provide a qualitative evaluation in Section~\ref{sec:eval}. To demonstrate the viability of our approach, we show that many UNIX programs, e.g.
\texttt{gzip} or a number of programs from the GNU \texttt{coreutils} suite can be compiled for \dios{} with no changes and subsequently analysed using an explicit-state model checker.

\hypertarget{related-work}{%
\subsection{Related Work}\label{related-work}}

Execution reproducibility is a widely studied problem. A number of tools capture \emph{provenance}, or history of the execution, by following and recording program's interactions with the environment,
later using this information to reproduce the recorded execution. For instance, ReproZip~\citep{chirigati13:reprozip} bundles the environment variables, files and library dependencies it observes so
that the executable can be run on a different system. Other programs exist, that instead capture the provenance in form of logs~\citep{shrinivas07:scarpe}, or sometimes more complex structures --
provenance graphs in case of ES3~\citep{frew08:automatic.capture}.

SCARPE~\citep{shrinivas07:scarpe} was developed for Java programs and captures I/O, user inputs and interactions with the database and the filesystem into a simple event log. The user has to state
which interactions to observe by annotating the individual classes that make up the program, since the instrumentation introduces substantial overhead and recording all interactions may generate a
considerable amount of data (for example, capturing a large portion of the database).

Another common approach to dealing with the complexity of interactions with the execution environment is \emph{mocking}~\citep{mackinnon01:endo.testing, mostafa14:empirical.study}: essentially
building small models of the parts of the environment that are relevant in the given test scenario. A \emph{mock object} is one step above a stub, which simply accepts and discards all requests. A
major downside of using mock objects in testing is that sufficiently modelling the environment requires a lot of effort: either the library only provides simple objects and users have to model the
system themselves, or the mock system is sophisticated, but the user has to learn a complex API.

Most testing frameworks for mainstream programming languages offer a degree of support for building mock objects, including mock objects which model interaction with the operating system. For instance
the \texttt{pytest} tool~\citep{krekel04:pytest} for Python allows the user to comfortably mock a database connection. A more complex example of mocking would be the filesystem support in
Pex~\citep{kong09:automated.testing}, a symbolic executor for programs targeting the .NET platform. \klee{} is a symbolic executor based on \llvm{} and targets C (and to some degree C++) programs with a
different approach to environment interaction. Instead of modelling the file system or other operating system services, it allows the program to directly interact with the host operating system,
optionally via a simple adaptation layer which provides a degree of isolation based on symbolic file models.

This latter approach, where system calls and even library calls are forwarded to the host operating system is also used in some runtime model checkers, most notably Inspect~\citep{yang08:inspec} and
CHESS~\citep{musuvathi08:findin.reprod}. However, those approaches, only work when the program interacts with the operating system in a way free from side effects, and when external changes in the
environment do not disturb verification. Finally, standard (offline) model checkers rarely support more than a handful of interfaces. The most widely supported is the \posix{} threading API, which is
modelled by tools such as Lazy-CSeq~\citep{inverso15:lazy.cseq} and its variants, by Impara~\citep{wachter13:verify} and by a few other tools.

\hypertarget{sec:platform}{%
\section{Platform Interface}\label{sec:platform}}

In this section, we will describe our expectations of the execution or verification platform and the low-level interface between this platform and our model operating system. We then break down the
interface into a small number of areas, each covering particular functionality.

\hypertarget{preliminaries}{%
\subsection{Preliminaries}\label{preliminaries}}

The underlying platform can have different characteristics. We are mainly interested in platforms or tools based on dynamic analysis, where the program is at least partially interpreted or executed,
often in isolation from the environment. If the platform itself isolates the system under test, many standard facilities like file system access become unavailable. In this case, the role of \dios{} is
to provide a substitute for the inaccessible host system.

If, on the other hand, the platform allows the program to access the host system, this easily leads to inconsistencies, where executions explored first can interfere with the state of the system
observed by executions explored later. For instance, files or directories might be left around, causing unexpected changes in the behaviour\footnote{If execution A creates a file and leaves it around,
  execution B might get derailed when it tries to create the same file, or might detect its presence and behave differently.} of the system under test. In cases like those, \dios{} can serve to insulate
such executions from each other. Under \dios{}, the program can observe the effects of its actions along a single execution path -- for instance, if the program creates a file, it will be able to open it
later. However, this file never becomes visible to another execution of the same program, regardless of the exploration order.

Unfortunately, not all facilities that operating systems provide to programs can be modelled entirely in terms of standard C. To the contrary, certain areas of high-level functionality that the
operating system is expected to implement strongly depend on low-level aspects of the underlying platform. Some of those are support for thread scheduling, process isolation, control flow constructs
such as \texttt{setjmp} and C++ exceptions, among others. We will discuss those in more detail in the following sections.

\hypertarget{program-memory}{%
\subsection{Program Memory}\label{program-memory}}

An important consideration when designing an operating system is the semantics of the memory subsystem of its execution platform. \dios{} is no exception: it needs to provide a high-level memory
management API to the application (both the C \texttt{malloc} interface and the C++ \texttt{new}/\texttt{delete} interface). In principle, a single flat array of memory is sufficient to implement all
the essential functionality. However, it lacks both in efficiency and in robustness. Ideally, the platform would provide a memory management API that manages individual memory objects which in turn
support an in-place resize operation. This makes operations more efficient by avoiding the need to make copies when extra memory is required, and the operating system logic simpler by avoiding a level
of indirection.

If the underlying platform is memory safe and if it provides a supervisor mode to protect access to certain registers or to a special memory location, the remainder of kernel isolation is implemented
by \dios{} itself, by withholding addresses of kernel objects from the user program. In this context, memory safety entails bound checks and an inability to overflow pointers from one memory object into
another.

\hypertarget{sec:stack}{%
\subsection{Execution Stack}\label{sec:stack}}

Information about active procedure calls and about the local data of each procedure are, on most platforms, stored in a special \emph{execution stack}. While the presence of such a stack is almost
universal, the actual representation of this stack is very platform-specific. On most platforms that we consider,\footnote{The main exception is \klee{}, where the execution stack is completely
  inaccessible to the program under test and only the virtual machine can access the information stored in it. See also Section~\ref{sec:klee}.} it is part of standard program memory and can be
directly accessed using standard memory operations. If both reading and modifications of the stack (or stacks) is possible, most of the operations that \dios{} needs to perform can be implemented without
special assistance from the platform itself. Those operations are:

\begin{itemize}
\tightlist
\item
  creation of a new execution stack is needed in two scenarios: isolation of the kernel stack from the user-space stack and creation of new tasks (threads, co-routines or other similar high-level
  constructs),
\item
  stack unwinding, where stack frames are traversed and removed from the stack during exception propagation or due to \texttt{setjmp}/\texttt{longjmp}.
\end{itemize}

Additionally, \dios{} needs a single operation that must be always provided by the underlying platform: it needs to be able to transfer control to a particular stack frame, whether within a single
execution stack (to implement non-local control flow) or to a different stack entirely (to implement task switching).

In some sense, this part of the platform support is the most complex and most tricky to implement. Fortunately, the features that rely on the above operations, or rather the modules which implement
those features, are all optional in \dios{}.

\hypertarget{auxiliary-interfaces}{%
\subsection{Auxiliary Interfaces}\label{auxiliary-interfaces}}

There are three other points of contact between \dios{} and the underlying platform. They are all optional or can be emulated using standard C features, but if available, \dios{} can use them to offer
additional facilities mainly aimed at software verification and testing with fault injection.

\hypertarget{indeterminate-values.}{%
\paragraph{Indeterminate values.}\label{indeterminate-values.}}

A few components in \dios{} use, or can be configured to use, values which are not a priori determined. The values are usually subject to constraints, but within those constraints, each possible value
will correspond to a particular interaction outcome. This facility is used for simulating interactions that depend on random chance (e.g.~thread scheduling, incidence of clock ticks relative to the
instruction stream), or where the user would prefer to not provide specific input values and instead rely on the verification or testing platform to explore the possibilities for them (e.g.~the
content of a particular file).

\hypertarget{nondeterministic-choice.}{%
\paragraph{Nondeterministic choice.}\label{nondeterministic-choice.}}

A special case of the above, where the selection is among a small number of discrete options. In those cases, a specific interface can give better user experience or better tool performance. If the
choice operator is not available but indeterminate values are, they can be used instead. Otherwise, the sequence of choices can be provided as an input by the user, or they can be selected randomly.
The choice operation is used for scheduling choices and for fault injection (e.g.~simulation of \texttt{malloc} failures).

\hypertarget{host-system-call-execution.}{%
\paragraph{Host system call execution.}\label{host-system-call-execution.}}

Most \posix{} operating systems provide an indirect system call facility, usually as the C function \texttt{syscall()}. If the platform makes this function accessible from within the system under test,
\dios{} can use it to allow real interactions between the user program and the host operating system to take place and to record and then replay such interactions in a reproducible manner.

\hypertarget{sec:supported}{%
\section{Supported Platforms}\label{sec:supported}}

In the previous section, we have described the target platform in generic, abstract terms. In this section, we describe 3 specific platforms which can execute \dios{} and how they fit with the above
abstract requirements.

\hypertarget{divm}{%
\subsection{\divm{}}\label{divm}}

\divm{}~\citep{rockai18:divm} is a verification-oriented virtual machine based on \llvm{}. A suite of tools based on \divm{} implement a number of software verification techniques, including explicit-state,
symbolic and abstraction-based model checking. \divm{} is the best supported of all the platforms, since it has been specifically designed to delegate responsibility for features to a model operating
system. All features available in \dios{} are fully supported on this platform.

In \divm{}, the functionality that is not accessible through standard C (or \llvm{}) constructs is provided via a set of \emph{hypercalls}. These hypercalls form the core of the platform interface in \dios{}
and whenever possible, ports to other platforms are encouraged to emulate the \divm{} hypercall interface using the available platform-native facilities.

\hypertarget{sec:klee}{%
\subsection{\klee{}}\label{sec:klee}}

\klee{}~\citep{cadar08:klee} is a symbolic executor based on \llvm{}, suitable both for automated test generation and for exhaustive exploration of bounded executions. Unlike \divm{}, \klee{} by default allows
the program under test to perform external calls (including calls to the host operating system), with no isolation between different execution branches. Additionally, such calls must be given concrete
arguments, since they are executed as native machine code (i.e.~not symbolically). However, if the program is linked to \dios{}, both these limitations are lifted: \dios{} code can be executed symbolically
like the rest of the program, and different execution branches are isolated from each other.

However, there is also a number of limitations when \klee{} is considered as a platform for \dios{}. The two most important are as follows:

\begin{enumerate}
\def\labelenumi{\arabic{enumi}.}
\tightlist
\item
  \klee{} does not currently support in-place resizing of memory objects. This is a design limitation and lifting it requires considerable changes. A workaround exists, but it is rather inefficient.
\item
  There is only one execution stack in \klee{} and there is no support for non-local control flow. This prevents \dios{} from offering threads, C++ exceptions and \texttt{setjmp} when executing in \klee{}.
\end{enumerate}

Additionally, there is no supervisor mode and hence no isolation between the kernel and the user program. However, in most cases, this is not a substantial problem. Non-deterministic choice is
available via indeterminate symbolic values and even though \klee{} can in principle provide access to host syscalls, we have not evaluated this functionality in conjunction with \dios{}. Finally, there are
a few minor issues that are, however, easily corrected:\footnote{A version of \klee{} with fixes for those problems is available online, along with other supplementary material, from
  \url{https://divine.fi.muni.cz/2019/dios/}.}

\begin{enumerate}
\def\labelenumi{\arabic{enumi}.}
\tightlist
\item
  \klee{} does not support the \texttt{va\_arg} \llvm{} instruction and relies on emulating platform-specific mechanisms instead, which are absent from \dios{},
\item
  it handles certain C functions specially, including the \texttt{malloc} family, the C++ \texttt{new} operator, the \texttt{errno} location and functions related to assertions and program
  termination; this interferes with the equivalent functionality provided by \dios{} \texttt{libc}, and finally
\item
  global constructors present in the program are unconditionally executed before the entry function; since \dios{} invokes constructors itself, this \klee{} behaviour also causes a conflict.
\end{enumerate}

\hypertarget{native-execution}{%
\subsection{Native Execution}\label{native-execution}}

\begin{figure}
\hypertarget{fig:native}{%
\centering
\includegraphics{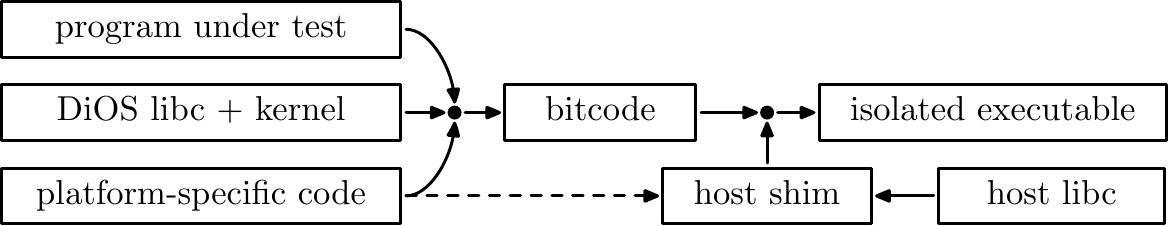}
\caption{Architecture of the native execution platform.}\label{fig:native}
}
\end{figure}

The third platform that we consider is native execution, i.e.~the \dios{} kernel is compiled into machine code, like a standard user-space program, to execute as a process of the host operating system.
This setup is useful in testing or in stateless model checking, where it can provide superior execution speed at the expense of reduced runtime safety. The user program still uses \dios{} \texttt{libc}
and the program runs in isolation from the host system. The platform-specific code in \dios{} uses a few hooks provided by a shim which calls through into the host operating system for certain services,
like the creation and switching of stacks. The design is illustrated in Figure~\ref{fig:native}.

Like in \klee{}, the native port of \dios{} does not have access to in-place resizing of memory objects, but it can be emulated slightly more efficiently using the \texttt{mmap} host system call. The native
port, however, does not suffer from the single stack limitations that \klee{} does: new stacks can be created using \texttt{mmap} calls and stack switching can be implemented using host \texttt{setjmp}
and \texttt{longjmp} functions.\footnote{The details of how this is done are quite technical, and are discussed in the online supplementary material at \url{https://divine.fi.muni.cz/2019/dios/}.} The
host stack unwinding code is directly used (the \divm{} platform code implements the same \texttt{libunwind} API that most \posix{} systems also use).

On the other hand, non-deterministic choice is not directly available. It can be simulated by using the \texttt{fork} host system call to split execution, but this does not scale to frequent choices,
such as those arising from scheduling decisions. In this case, a random or an externally supplied list of outcomes are the only options.

\hypertarget{sec:dios}{%
\section{Design and Architecture}\label{sec:dios}}

\begin{figure}
\hypertarget{fig:arch}{%
\centering
\includegraphics{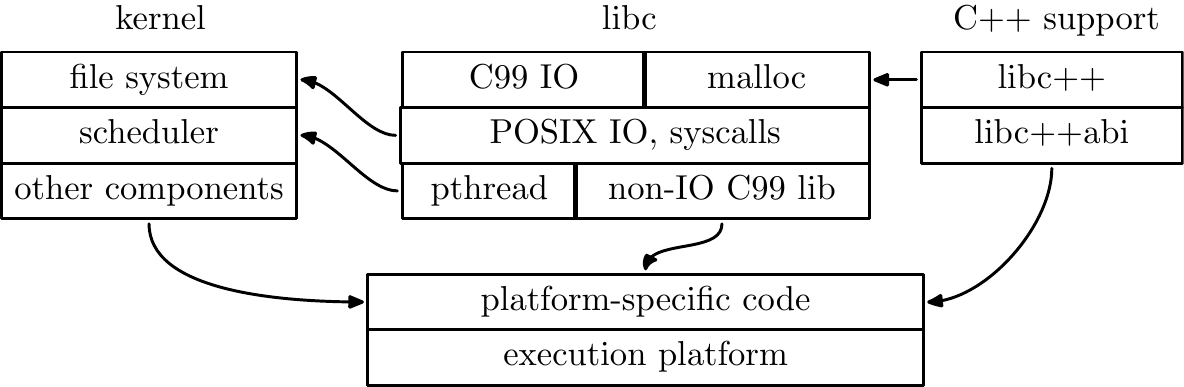}
\caption{The architecture of \dios{}.}\label{fig:arch}
}
\end{figure}

This section outlines the structure of the \dios{} kernel and userspace, their components and the interfaces between them. We also discuss how the kernel interacts with the underlying platform and the
user-space libraries stacked above it. A high-level overview of the system is shown in Figure~\ref{fig:arch}. The kernel and the user-mode parts of the system under test can be combined using
different methods; even though they can be linked into a single executable image, this is not a requirement, and the kernel can operate in a separate address space.

Like with traditional operating systems, kernel memory is inaccessible to the program and libraries executing in user-mode. In \dios{}, this protection is optional, since not all platforms provide
supervisor mode or sufficient memory safety; however, it does not depend on address space separation between the kernel and the user mode.

\hypertarget{sec:components}{%
\subsection{Kernel Components}\label{sec:components}}

The decomposition of the kernel to a number of components serves multiple goals: first is resource conservation -- some components have non-negligible memory overhead even when they are not actively
used. This may be because they need to store auxiliary data along with each thread or process, and the underlying verification tool then needs to track this data throughout the execution or throughout
the entire state space. The second is improved portability to platforms which do not provide sufficient support for some of the components, for instance thread scheduling. Finally, it allows \dios{} to
be reconfigured to serve in new contexts by adding a new module and combining it with existing code.

The components of the kernel are organised as a stack, where upper components can use services of the components below them. While this might appear to be a significant limitation, in practice this
has not posed substantial challenges, and the stack-organised design is both efficient and simple. A number of pre-made components are available, some in multiple alternative implementations:

\hypertarget{task-scheduling-and-process-management.}{%
\paragraph{Task scheduling and process management.}\label{task-scheduling-and-process-management.}}

There are 4 scheduler implementations: the simplest is a \emph{null} scheduler, which only allows a single task and does not support any form of task switching. This scheduler is used on \klee{}. Second
is a synchronous scheduler suitable for executing software models of hardware devices. The remaining two schedulers both implement asynchronous, thread-based parallelism. One is designed for
verification of safety properties of parallel programs, while the other includes a fairness provision and is therefore more suitable for verification of liveness properties.

In addition to the scheduler, there is an optional process management component. It is currently only available on the \divm{} platform, since it heavily relies on operations which are not available
elsewhere. It implements the \texttt{fork} system call and requires one of the two asynchronous schedulers.

\hypertarget{posix-system-calls.}{%
\paragraph{\posix{} System Calls.}\label{posix-system-calls.}}

While a few process-related system calls are implemented in the components already mentioned, the vast majority is not. By far the largest coherent group of system calls deals with files, directories,
pipes and sockets, where the unifying concept is file descriptors. A memory-backed filesystem module implements those system calls by default.

A smaller group of system calls relate to time and clocks and those are implemented in a separate component which simulates a system clock. The specific simulation mode is configurable and can use
either indeterminate values to shift the clock every time it is observed or a simpler variant where ticks of fixed length are performed based on the outcome of a nondeterministic choice.

The system calls covered by the filesystem and clock modules can be alternately provided by a \emph{proxy} module, which forwards the calls to the host operating system or by a \emph{replay} module
which replays traces captured by the \emph{proxy} module.

\hypertarget{auxiliary-modules.}{%
\paragraph{Auxiliary modules.}\label{auxiliary-modules.}}

There is a small number of additional modules which do not directly expose functionality to the user program. Instead, they fill in support roles within the system. The two notable examples are the
\emph{fault handler} and the \emph{system call stub} component.

The fault handler takes care of responding to error conditions indicated by the underlying platform. It is optional, since not all platforms can report problems to the system under test. If present,
the component allows the user to configure which problems should be reported as counterexamples and which should be ignored. The rest of \dios{} also uses this component to report problems detected by
the operating system itself, e.g.~the \texttt{libc} uses it to flag assertion failures.

The stub component supplies fallback implementations of all system calls known to \dios{}. This component is always at the bottom of the kernel configuration stack -- if any other component in the active
configuration implements a particular system call, that implementation is used. Otherwise, the fallback is called and raises a runtime error, indicating that the system call is not supported.

\hypertarget{sec:threads}{%
\subsection{Thread Support}\label{sec:threads}}

One of the innovative features of \dios{} is that it implements the \posix{} threading API using a very simple platform interface. Essentially, the asynchronous schedulers in \dios{} provide an illusion of
thread-based parallelism to the program under test, but only use primitives associated with coroutines -- creation and switching of execution stacks (cf.~Section~\ref{sec:stack}).

However, an additional external component is required: both user and library code needs to be instrumented with \emph{interrupt points}, which allow thread preemption to take place. Where to insert
them can be either decided statically (which is sufficient for small programs) or dynamically, allowing the state space to be reduced using more sophisticated techniques.\footnote{In
  \divine{}~\citep{baranova17:model.checkin}, a model checker based on \divm{}, interrupt points are dynamically enabled when the executing thread performs a visible action. Thread identification is
  supplied by the scheduler in \dios{} using a platform-specific (hypercall) interface.} The implementation of the interrupt point is, however, supplied by \dios{}: only the insertion of the function call
is done externally.

The scheduler itself provides a very minimal internal interface -- the remainder of thread support is implemented in user-space libraries (partly \texttt{libc} and partly \texttt{libpthread}, as is
common on standard \posix{} operating systems). Even though the implementation is not complete (some of the rarely-used functions are stubbed out), all major areas are well supported: thread creation and
cancellation, mutual exclusion, condition variables, barriers, reader-writer locks, interaction with \texttt{fork}, and thread-local storage are all covered. Additionally, both C11 and C++11 thread
APIs are implemented in terms of the \texttt{pthread} interface.

\hypertarget{sec:syscalls}{%
\subsection{System Calls}\label{sec:syscalls}}

The system call interface of \dios{} is based on the ideas used in \emph{fast system call} implementations on modern processors.\footnote{For instance, on contemporary \texttt{x86-64} processors, this
  interface is available via the \texttt{syscall} and \texttt{sysret} instructions.} A major advantage of this approach is that system calls can be performed using standard procedure calls on
platforms which do not implement supervisor mode.

The list of system calls available in \dios{} is fixed:\footnote{The list of system calls is only fixed relative to the host operating system. To allow the system call proxy component to function
  properly, the list needs to match what is available on the host. For instance, \texttt{creat}, \texttt{uname} or \texttt{fdatasync} are system calls on Linux but standard \texttt{libc} functions on
  OpenBSD.} in addition to the kernel-side implementation, which may or may not be available depending on the active configuration, each system call has an associated user-space C function, which is
declared in one of the public header files and implemented in \texttt{libc}.

The available system calls cover thread management, sufficient to implement the \texttt{pthread} interface (the system calls themselves are not standardised by \posix{}), the \texttt{fork} system call,
\texttt{kill} and other signal-related calls, various process and process group management calls (\texttt{getpid}, \texttt{getsid}, \texttt{setsid}, \texttt{wait} and so on). Notably, \texttt{exec} is
currently not implemented and it is not clear whether adding it is feasible on any of the platforms. The thread- and process- related functionality was described in more detail in
Section~\ref{sec:threads}.

Another large group of system calls cover files and networking, including the standard suite of \posix{} calls for opening and closing files, reading and writing data, creating soft and hard links. This
includes the \texttt{*at} family introduced in \posix{}.1 which allows thread-safe use of relative paths. The standard BSD socket API is also implemented, allowing threads or processes of the program
under test to use sockets for communication. Finally, there are system calls for reading (\texttt{clock\_gettime}, \texttt{gettimeofday}) and setting clocks (\texttt{clock\_settime},
\texttt{settimeofday}).

\hypertarget{sec:libc}{%
\subsection{The C Library}\label{sec:libc}}

\dios{} comes with a complete ISO C99 standard library and the C11 thread API. The functionality of the C library can be broken down into the following categories:

\begin{itemize}
\tightlist
\item
  Input and output. The functionality required by ISO C is implemented in terms of the \posix{} file system API. Number conversion (for formatted input and output) is platform independent and comes from
  \texttt{pdclib}.
\item
  The string manipulation and character classification routines are completely system-independent. The implementations were also taken from \texttt{pdclib}.
\item
  Memory allocation: new memory needs to be obtained in a platform-dependent way. Optionally, memory allocation failures can be simulated using a non-deterministic choice operator. The library
  provides the standard assortment of functions: \texttt{malloc}, \texttt{calloc}, \texttt{realloc} and \texttt{free}.
\item
  Support for \texttt{errno}: this variable holds the code of the most recent error encountered in an API call. On platforms with threads (like \dios{}), \texttt{errno} is thread-local.
\item
  Multibyte strings: conversions of Unicode character sequences to and from UTF-8 is supported.
\item
  Time-related functions: time and date formatting (\texttt{asctime}) is supported, as is obtaining and manipulating wall time. Interval timers are currently not simulated, although the relevant
  functions are present as simple stubs.
\item
  Non-local jumps. The \texttt{setjmp} and \texttt{longjmp} functions are supported on \divm{} and native execution, but not in \klee{}.
\end{itemize}

In addition to ISO C99, there are a few extensions (not directly related to the system call interface) mandated by \posix{} for the C library:

\begin{itemize}
\tightlist
\item
  Regular expressions. The \dios{} \texttt{libc} supports the standard \texttt{regcomp} \& \texttt{regexec} APIs, with implementation based on the TRE library.
\item
  Locale support: A very minimal support for \posix{} internationalisation and localisation APIs is present. The support is sufficient to run programs which initialise the subsystem.
\item
  Parsing command line options: the \texttt{getopt} and \texttt{getopt\_long} functions exist to make it easy for programs to parse standard UNIX-style command switches. \dios{} contains an
  implementation derived from the OpenBSD code base.
\end{itemize}

Finally, C99 mandates a long list of functions for floating point math, including trigonometry, hyperbolic functions and so on. A complete set of those functions is provided by \dios{} via its
\texttt{libm} implementation, based on the OpenBSD version of this library.

\hypertarget{c-support-libraries}{%
\subsection{C++ Support Libraries}\label{c-support-libraries}}

\dios{} includes support for C++ programs, up to and including the C++17 standard. This support is based on the \texttt{libc++abi} and \texttt{libc++} open-source libraries maintained by the \llvm{}
project. The versions bundled with \dios{} contain only very minor modifications relative to upstream, mainly intended to reduce program size and memory use in verification scenarios.

Notably, the exception support code in \texttt{libc++abi} is unmodified and works both in \divm{} and when \dios{} is executing natively as a process of the host operating system. This is because
\texttt{libc++abi} uses the \texttt{libunwind} library to implement exceptions. When \dios{} runs natively, the host version of \texttt{libunwind} is used, the same as with \texttt{setjmp}. When
executing in \divm{}, \dios{} supplies its own implementation of the \texttt{libunwind} API, as described in~\citep{still17:using.off}.

\hypertarget{sec:abi}{%
\subsection{Binary Compatibility}\label{sec:abi}}

When dealing with verification of real-world software, the exact layout of data structures becomes important, mainly because we would like to generate native code from verified bitcode files (when
using either \klee{} or \divm{}). To this end, the layouts of relevant data structures and values of relevant constants are automatically extracted from the host operating system\footnote{This extraction is
  performed at \dios{} build time, using \texttt{hostabi.pl}, which is part of the \dios{} source distribution. The technical details are discussed in the online supplementary material.} and then used in
the \dios{} \texttt{libc}. As a result, the native code generated from the verified bitcode can be linked to host libraries and executed as usual. The effectiveness of this approach is evaluated in
Section~\ref{sec:abieval}.

\begin{figure}
\hypertarget{fig:host}{%
\centering
\includegraphics{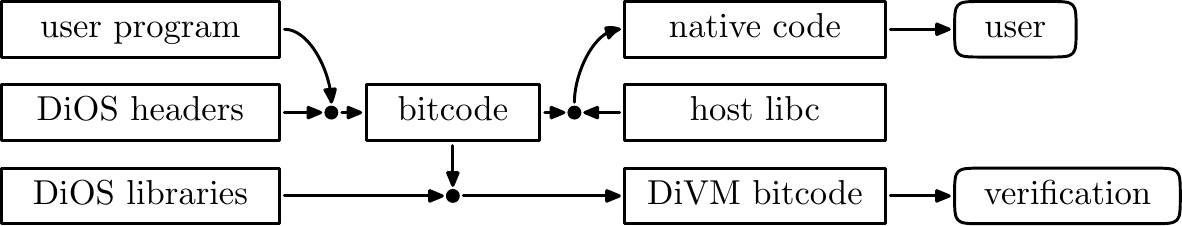}
\caption{Building verified executables with \dios{}.}\label{fig:host}
}
\end{figure}

\hypertarget{sec:eval}{%
\section{Evaluation}\label{sec:eval}}

We have tested \dios{} in a number of scenarios, to ensure that it meets the goals that we describe in Section~\ref{sec:goals}. The first goal -- modularity -- is hard to quantify in isolation, but it
was of considerable help in adapting \dios{} for different use cases. We have used \dios{} with success in explicit-state model checking of parallel programs~\citep{baranova17:model.checkin}, symbolic
verification of both parallel and sequential programs~\citep{lauko19:extend.divine}, for verification of liveness (LTL) properties of synchronous C code synthesized from Simulink diagrams, and for
runtime verification of safety properties of software~\citep{kejstova17:from.model}. \dios{} has also been used for recording, replaying and fuzzing system call
traces~\citep{kejstova19:model.checking.system}.

\hypertarget{verification-with-divm}{%
\subsection{Verification with \divm{}}\label{verification-with-divm}}

In this paper, we report on 3 sets of tests that we performed particularly to evaluate \dios{}. The first is a set of approximately 2200 test programs\footnote{All test programs are available online at
  \url{http://divine.fi.muni.cz/2019/dios/}, including scripts to reproduce the results reported in this and in the following sections.} which cover various aspects of the entire verification
platform. Each of them was executed in \dios{} running on top of \divm{} and checked for a number of safety criteria: lack of memory errors, use of uninitialized variables, assertion violations, deadlocks
and arithmetic errors. In the case of parallel programs (about 400 in total), all possible schedules were explored. Additionally, approximately 700 of the test programs depend on one or more input
values (possibly subject to constraints), in which case symbolic methods or abstraction are used to cover all feasible paths through the program. The tests were performed on two host operating
systems: Linux 4.19 with \texttt{glibc} 2.29 and on OpenBSD 6.5, with no observed differences in behaviour.

Majority (1300) of the programs are written in C, the remainder in C++, while a third of them (700) were taken from the SV-COMP~\citep{beyer16:reliab.reprod} benchmark suite. Roughly half of the
programs contain a safety violation, the location of which is annotated in the source code. The results of the automated analysis are in each case compared against the annotations and no mismatches
were found in the set.

\hypertarget{portability}{%
\subsection{Portability}\label{portability}}

To evaluate the remaining ports of \dios{}, we have taken a small subset (370 programs, or 17\%) of the entire test suite and executed the programs on the other two platforms currently supported by \dios{}.
The subset was selected to fall within the constraints imposed by the limitations of our \klee{} port -- in particular, lack of support for threads and for C++ exceptions. We have focused on filesystem
and socket support (50 programs) and exercising the standard C and C++ libraries shipped with \dios{}. The test cases have all completed successfully, and \klee{} has identified all the annotated safety
violations in these programs.

\hypertarget{sec:abieval}{%
\subsection{API and ABI Coverage and Compatibility}\label{sec:abieval}}

Finally to evaluate our third goal, we have compiled a number of real-world programs against \dios{} headers and libraries and manually checked that they behave as expected when executed in \dios{} running
on \divm{}, fully isolated from the host operating system. The compilation process itself exercises source-level (API) compatibility with the host operating system.

We have additionally generated native code from the bitcode that resulted from the compilation using \dios{} header files (see Figure~\ref{fig:host}) and which we confirmed to work with \dios{} libraries.
We then linked the resulting machine code with the \texttt{libc} of the host operating system (\texttt{glibc} 2.29 in this case). We have checked that the resulting executable program also behaves as
expected, confirming a high degree of binary compatibility with the host operating system. The programs we have used in this test were the following (all come from the GNU software collection):

\begin{itemize}
\tightlist
\item
  \texttt{coreutils} 8.30, a collection of 107 basic UNIX utilities, out of which 100 compiled successfully (we have tested a random selection of those),
\item
  \texttt{diffutils} 3.7, programs for computing differences between text files and applying the resulting patches -- the diffing programs compiled and \texttt{diff3} was checked to work correctly,
  while the \texttt{patch} program failed to build due to lack of \texttt{exec} support on \dios{},
\item
  \texttt{sed} 4.7 builds and works as expected,
\item
  \texttt{make} 4.2 builds and can parse makefiles, but it cannot execute any rules due to lack of \texttt{exec} support,
\item
  the \texttt{wget} download program failed to build due to lack of \texttt{gethostbyname} support, the cryptographic library \texttt{nettle} failed due to deficiencies in our compiler driver and
  \texttt{mtools} failed due to missing \texttt{langinfo.h} support.
\end{itemize}

\hypertarget{sec:conclusions}{%
\section{Conclusions \& Future Work}\label{sec:conclusions}}

We have presented \dios{}, a \posix{}-compatible operating system designed to offer reproducible execution, with special focus on applications in program verification. The larger goal of verifying
unmodified, real-world programs requires the cooperation of many components, and a model of the operating system is an important piece of the puzzle. As the case studies show, the proposed approach is
a viable way forward. Just as importantly, the design goals have been fulfilled: we have shown that \dios{} can be successfully ported to rather dissimilar platforms, and that its various components can
be disabled or replaced with ease.

Implementation-wise, there are two important future directions: further extending the coverage and compatibility of \dios{} with real operating systems, and improving support for different execution and
software verification platforms. In terms of design challenges, the current model of memory management for multi-process systems is suboptimal and there are currently no platforms on which the
\texttt{exec} family of system calls could be satisfactorily implemented. We would like to rectify both shortcomings in the future.

\bibliography{common}

\end{document}